\documentclass[final,3p,twocolumn,fleqn]{elsarticle}

\usepackage{hyperref}
\RequirePackage{graphicx}
\RequirePackage{flushend}
\RequirePackage{xparse}
\usepackage{physics}

\begin{document}

\begin{frontmatter}

\title{An analytical approach to the CMB Polarization in a Spatially Closed background}

\author{Pedram Niazy}
\ead{pedram.niazi@modares.ac.ir}

\author{Amir H. Abbassi}
\ead{ahabbasi@modares.ac.ir}

\address{Department of Physics, School of Science, Tarbiat Modares University. P.O. Box 14155-4838, Tehran, Iran}

\begin{abstract}
The scalar mode polarization of the cosmic microwave background is derived in a spatially closed universe from the Boltzmann equation using the line of sight integral method. The EE and TE multipole coefficients have been extracted \emph{analytically} by considering some tolerable approximations such as considering the evolution of perturbation hydrodynamically and sudden transition from opacity to transparency at the time of last scattering. As the major advantage of analytic expressions, $C^S_{EE,\ell}$ and $C^S_{TE,\ell}$ explicitly show the dependencies on baryon density $\Omega_B$, matter density $\Omega_M$, curvature $\Omega_K$, primordial spectral index $n_s$, primordial power spectrum amplitude $A_s$, Optical depth $\tau_{reion}$, recombination width $\sigma_t$ and recombination time $t_L$. Using a realistic set of cosmological parameters taken from a fit to data from Planck, the closed universe EE and TE power spectrums in the scalar mode are compared with numerical results from the CAMB code and also latest observational data. The analytic results agree with the numerical ones on the big and moderate scales. The peak positions are in good agreement with the numerical result on these scales while the peak heights agree with that to within $20\%$ due to the approximations have been considered for these derivations. Also, several interesting properties of CMB polarization are revealed by the analytic spectra.\\
\end{abstract}


\end{frontmatter}

%\linenumbers
\setlength{\mathindent}{0pt}
%%%%%%%%%%%%%%%%%%%%%%%%%%%%%%%%%%%%%%%%%%%%%%%%%%%%%%%%%%%%%%%%%%%%%%%%%%%%%
%Section 1
\section{Introduction}\label{601}
The temperature and polarization of the cosmic microwave background radiation (CMB) considered as an essential cosmological probe. The cosmological parameters of the standard big bang model can be considerably constrained by comparing the predictions of theoretical cosmological models with the data on the CMB by observation, such as WMAP and Planck. Although there are several numerical works in this field such as CAMB  \cite{r1,r2,r3} and CMBFAST \cite{r4}, analytical studies of the anisotropy and polarization give us a great insight into the problem by understanding how various underlying physical effects give rise to specific observational behavior.\\
There are several works in the field that extracted an analytical expression for the $C_{XX',\ell}$ in a flat universe such as Refs. \cite{r5,r6,r7,r8,r9,r10,r11} where all analytical spectra have been extracted by considering the tensor perturbation as a source. Ref. \cite{r18} also gave a unifying framework for all spectra in both tensor and scalar modes.\\
In our previous work \cite{r28} we performed a detail analytic calculation of the scalar mode (in synchronous gauge) temperature power spectrum $C^S_{TT,\ell}$ in a spatially closed background and now we continue that way to extract analytical expressions for EE and TE polarization multipole coefficients. Like what we have done for temperature power spectrum derivation, we will apply some of the results and techniques developed in the study of the CMB anisotropies in a flat spatial geometry by extending them into the closed geometry case. By employing some tolerable approximations, we extract simple \emph{analytic} formulas for the EE and TE polarization multipole coefficients which provide a transparent information about the dependencies of the CMB spectra on cosmological parameters.\\

In the following section, we give a brief overview of the perturbation theory in a spatially closed universe and then, by introducing the Boltzmann equation for the photons in the spatially closed background we review the procedures we have used in our previous work to extract the CMB temperature fluctuations and multipole moment $a^S_{T,\ell m}$. In section \ref{603}, we present a general formula for the polarization of the CMB in a spatially closed background and derive the polarization multipole moment $a^S_{E,\ell m}$. In section \ref{604}, we extract analytic formulas for the polarization multipole coefficients $C^S_{EE,\ell}$ and $C^S_{TE,\ell}$ by considering some approximations like a sudden transition from opacity to transparency at the time of last scattering and considering the evolution of perturbation hydrodynamically. In section \ref{605}, we plot the EE and TE power spectra extracted in section \ref{604} using a realistic set of cosmological parameters and compare them with the numerical results from CAMB and also the curve from latest observational data (Planck 2015). Several interesting properties of CMB polarization are revealed in analytic expression along with the power spectrum dependence on cosmological parameters in this section. We conclude the article by a brief review which remarks the main results of this paper.
%%%%%%%%%%%%%%%%%%%%%%%%%%%%%%%%%%%%%%%%%%%%%%%%%%%%%%%%%%%%%%%%%%%%%%%%%%%%%
%Section 2
\section{Temperature fluctuations from the Boltzmann equation formalism; a short review}\label{602}
In this section we give a brief overview of the perturbation theory in a spatially closed universe and then review the procedures we have used in our previous work \cite{r28} to extract the scalar mode temperature fluctuation from the Boltzmann equation.\\

The theory of the linear perturbations is an important part of the modern cosmology which explains CMB anisotropies and the origin of structure formation.\\
We assume that during most of the history of the universe all departures from homogeneity and isotropy have been small so that they can be treated as first-order perturbations. We will take the unperturbed matric to have the Robertson-Walker form. The total perturbed metric is then
\begin{equation}\label{1}
  g_{\mu\nu}=\overline{g}_{\mu\nu}+h_{\mu\nu}
\end{equation}
where $\overline{g}_{\mu\nu}$ is the unperturbed metric FLRW metric which in the comoving spherical polar coordinates can be written as
\begin{align*}
  \overline{g}_{00}=&-1\\
  \overline{g}_{rr}=&{\frac{a^2 (t)}{1-Kr^2}\quad
  \overline{g}_{\theta\theta}=a^2(t)r^2\quad
  \overline{g}_{\varphi\varphi}=a^2(t)r^2\sin^2\theta}
\end{align*}
and $h_{\mu\nu}$ is a small perturbation.
Perturbation in the metric leads to perturbation in the Ricci and energy-momentum tensor. We can decompose the metric perturbation and energy-momentum tensors into the scalar, vector and tensor modes from their transformation properties under spatial rotations and derive the field equations accordingly \cite{r19}.\\
Decomposition into the scalar, vector and tensor modes of the metric perturbation and energy-momentum tensor would be as follows:
\begin{align*}
  h_{00} =& -E \\
  h_{i0} =& a\,(\nabla_i F+G_i)\\
  h_{ij} =& a^2\,(A\,\widetilde{g}_{ij}+H_{ij} B+\nabla_i C_j+\nabla_j C_i+D_{ij})\\
  \delta T_{00} =& -\overline{\rho}\,h_{00}+\delta \rho\\
  \delta T_{i0} =& \overline{p}\,h_{i0}-(\overline{\rho} + \overline{p} )(\nabla_i \delta u+\delta u_i^V)\\
  \delta T_{ij} =& \overline{p}\, h_{ij}+a^2 (\widetilde{g}_{ij}\delta p+H_{ij} \Pi^S+\nabla_i \Pi_j^V+\nabla_j \Pi_i^V+\Pi_{ij}^T).
\end{align*}
where $\nabla_i$ is the covariant derivative with respect to the spatial unperturbed metric $\widetilde{g}_{ij}(=a^{-2}\overline{g}_{ij})$ and $H_{ij}=\nabla_i \nabla_j$ is the \emph{covariant Hessian operator}. All the perturbations $A\,$, $B\,$, $E\,$, $F\,$, $C_i\,$, $G_i\,$ and $D_{ij}$ are functions of $\textbf{x}$ and $t$ which satisfy
\begin{align*}
  &\nabla^i C_i=\nabla^i G_i=0\\
  &\widetilde{g}^{ij}D_{ij}=0\quad \nabla^i D_{ij}=0\quad D_{ij}=D_{ji}
\end{align*}
On the other hand, all above perturbative quantities have been considered as random fields on $S^3(\alpha)$ (a 3-sphere of radius $\alpha$), because they are defined on a homogeneous and isotropic space \cite{r20,r21}. So they can be described by their Fourier transformation. There are many different Fourier transform convention, however here we are going to expand each mode of the perturbation fields in terms of the corresponding eigenfunctions of the \emph{Laplace-Beltrami} operator. This operator reduces to the ordinary Laplacian in a flat background. In pseudo-spherical coordinates with the line element
\begin{equation}\label{2}
  ds^2=\alpha^2\,(\textrm d\chi^2+\sin^2 \chi\, \textrm d\theta^2 +\sin^2 \chi\, \sin^2 \theta \,\textrm d\varphi^2  )
\end{equation}
one gets the following eigenvalues and eigenfunctions for the Laplace-Beltrami operator:
\begin{align*}
  &\nabla^2\Phi =  -k_n^2 \Phi \qquad \nabla^2=\widetilde{g}_{ij}H_{ij}=\widetilde{g}_{ij}\nabla_i\nabla_j\\
  &\Phi = \mathcal{Y}_{n\ell m} (\chi,\theta,\varphi)=\Pi_{n\ell} (\chi)\, Y_{\ell m} (\theta,\varphi)\\
  &k_n^2 =  \frac{n^2-1}{\alpha^2}\qquad n=1,2,\cdots
\end{align*}
where $\Pi_{n\ell}(\chi)$ is the hyperspherical Bessel function satisfying the following equation
\begin{align}\label{3}
  \dv[2]{\Pi_{n\ell}(\chi)}{\chi}+&2\cot\chi\dv{\Pi_{n\ell}(\chi)}{\chi}\nonumber\\
  +&\left[(n^2-1)-\frac{\ell(\ell+1)}{\sin^2\chi}\right]{\Pi_{n\ell}(\chi)}=0
\end{align}
In a flat background, the hyperspherical Bessel function reduces to the ordinary spherical Bessel function $j_\ell (\nu\chi)$.
Also we introduce the generalized wave number in closed space $q_n$ as
\begin{align*}
    q_n = \sqrt{k_n^2 + \frac{1}{\alpha^2}} = \frac{n}{\alpha}
\end{align*}
We can expand the scalar perturbative quantity $A(\textbf{x},t)$ in terms of \emph{Laplace-Beltrami} operator eigenfunctions as below:
\begin{equation}\label{4}
 A(\textbf{x},t)=\sum_{n\ell m}A_{n\ell m}(t)\,\mathcal{Y}_{n\ell m}(\chi,\theta,\varphi)
\end{equation}
This is the initial conditions that depend on the direction, not the perturbation itself, so a perturbation can be shown by a time-dependent normal mode $A_n (t)$ with an overall normalization factor $\alpha_{lm}$.
$A_{n\ell m}(t)$ just like $A(\textbf{x},t)$ is a scalar random field and one of the simplest statistics for it is the two-point covariant function denoted by $\langle A_{n\ell m} A^\ast_{n'\ell'm'}\rangle$. Here $\langle\quad\rangle$ means the ensemble average which equals the spatial average according to the \emph{ergodic} theorem.
The homogeneity and isotropy imply that
\begin{equation*}
  \langle\alpha_{\ell m}\,\alpha_{\ell' m'}^\ast\rangle=\delta_{\ell\ell'}\,\delta_{mm'}\quad\langle A_n(t)\,A_{n'}^\ast(t)\rangle=A_n^2(t)\,\delta_{nn'}
\end{equation*}
so the two-point covariant function of $A_{n\ell m}(t)$ is
\begin{align*}
  \langle A_{n\ell m} A^\ast_{n'\ell'm'}\rangle
   &= \langle\alpha_{\ell m}\,A_n(t)\,\alpha_{\ell m}^\ast\,A_{n'}^\ast(t)\rangle\\
   &=\langle\alpha_{\ell m}\,\alpha_{\ell' m'}^\ast\rangle\langle A_n(t)\,A_{n'}^\ast(t)\rangle\\
   &=A_n^2(t)\,\delta_{nn'}\,\delta_{\ell\ell'}\,\delta_{mm'}
\end{align*}
and for any scalar random field $A$, we will have
\begin{equation}\label{5}
 A(\textbf{x},t)=\sum_{n\ell m}\alpha_{\ell m}\,A_n(t)\,\mathcal{Y}_{n\ell m}(\chi,\theta,\varphi)
\end{equation}
Also, we can decompose an arbitrary tensor random field using the scalar eigenvalues of the Laplace-Beltrami operator and their covariant derivatives as follows \cite{r22,r23}
\begin{align}\label{6}
  A^{ij}(\textbf{x},t)&=\sum_{n\ell m}\alpha_{\ell m}\bigg[\frac{1}{3}A_{nT}(t)\,\tilde{g}^{ij}\,\mathcal{Y}_{nlm}(\chi,\theta,\varphi)\nonumber\\
  &+A_{nTL}(t)\left(k_n^{-2}H^{ij}\mathcal{Y}_{nlm}(\chi,\theta,\varphi)\right) \bigg],
\end{align}
where $A_{nT}(t)$ and $A_{nTL}(t)$ are the trace and traceless parts of the tensor $A^{ij}$ respectively.

Now, we move on the derivation of the CMB temperature fluctuations from the Boltzmann equation.\\
The Boltzmann equation that governs the evolution of the distribution of photons in phase space, can be written as
\begin{align*}
  \pdv{n_\gamma^{ij}}{t}&+\pdv{n_\gamma^{ij}}{x^k}\frac{p^k}{p^0}+\pdv{n_\gamma^{ij}}{p_k}\frac{p^lp^m}{2p^0}\pdv{g_{lm}}{x^k}\nonumber\\
   &+\left(\Gamma^i_{k\lambda}-\frac{p_i}{p^0}\Gamma^0_{k\lambda}\right)\frac{p^\lambda}{p^0}n^{kj}_\gamma\nonumber\\
   &+\left(\Gamma^j_{k\lambda}-\frac{p_j}{p^0}\Gamma^0_{k\lambda}\right)\frac{p^\lambda}{p^0}n^{ki}_\gamma=C^{ij}
\end{align*}
where $n_\gamma^{ij}$ is the number density matrix for photons and $C^{ij}$ is a term representing the effect of photon scattering.
Let us introduce the dimensionless intensity matrix as follows \cite{r24}
\begin{equation*}
  a^4(t)\overline{\rho}_\gamma(t)J^{ij}(\textbf{x},\hat{p},t)\equiv a^2(t)\int_{0}^{\infty}\delta n_\gamma^{ij}(\textbf{x},p\hat{p},t)4\pi\,p^3 \dd p
\end{equation*}
One can gets the Boltzmann equation in terms of $J^{ij}(\textbf{x},\hat{p},t)$ matrix as
\begin{align}\label{260}
 &\pdv{t}J^{ij}(\textbf{x},\hat{p},t)
 +\frac{\hat{p}^k}{a(t)}\pdv{x^k}J^{ij}(\textbf{x},\hat{p},t)\nonumber\\
 +&\frac{\hat{p}^s}{a(t)}\left(\widetilde{\Gamma}^i_{ks}J^{kj}(\textbf{x},\hat{p},t)+\widetilde{\Gamma}^j_{ks}J^{ki}(\textbf{x},\hat{p},t)\right)\nonumber\\
 +&\hat{p}^s\hat{p}^t\pdv{t}\left(a^{-2}\delta g_{st}\right)\left(\tilde{g}^{ij}-\hat{p}^i\hat{p}^j\right)\nonumber\\
 -&2\,\frac{\hat{p}^k\hat{p}^s\hat{p}^t}{a(t)}\partial_k\,\tilde{g}_{st}J^{ij}(\textbf{x},\hat{p},t)\nonumber\\
 =&-\omega_c(t)J^{ij}(\textbf{x},\hat{p},t)\nonumber\\
 +&\frac{2\,\omega_c(t)}{a(t)}\tilde{g}^{kl}\hat{p}_l\,\delta u_k(\textbf{x},t)\left(\tilde{g}^{ij}-\hat{p}^i\hat{p}^j\right)\nonumber\\
 +&\frac{3\,\omega_c(t)}{8\pi}\int \dd[2]{\hat{p}_1}\sqrt{Det\,\tilde{g}^{ij}}\bigg[J^{ij}(\textbf{x},\hat{p}_1,t)\nonumber\\
 -&\tilde{g}^{ik}\hat{p}_k\hat{p}_lJ^{jl}(\textbf{x},\hat{p}_1,t)-\tilde{g}^{jk}\hat{p}_k\hat{p}_lJ^{il}(\textbf{x},\hat{p}_1,t)\nonumber\\
 +&\tilde{g}^{ik}\tilde{g}^{jl}\hat{p}_k\hat{p}_l\hat{p}_m\hat{p}_nJ^{mn}(\textbf{x},\hat{p}_1,t)\bigg]
\end{align}
where $\omega_c(t)$ is the collision rate of a photon with electrons in the baryonic plasma and $\delta u_l(\textbf{x},t)$ is the peculiar velocity of the baryonic plasma. The term containing $\delta u_l (\textbf{x},t)$ will be added to the Boltzmann equation in scalar or vector modes.\\
Now using the perturbation theory in a spatially closed universe, we expand the metric perturbation, $J^{ij} (\textbf{x},\hat{p},t)$ and $\delta u_l (\textbf{x},t)$ in terms of the eigenfunctions of Laplace-Beltrami operator. The metric perturbation in scalar mode can be written as
\begin{equation*}
  a^{-2}(t)\delta g_{st}=A\,\tilde{g}_{st}+H_{st}\,B
\end{equation*}
where for the perturbative quantities $A$ and $B$ we can write
\begin{align*}
 A(\textbf{x},t)&=\sum_{n\ell m}\alpha_{\ell m}A_n(t)\mathcal{Y
 }_{n\ell m}(\chi,\theta,\phi)\\
 B(\textbf{x},t)&=\sum_{n\ell m}\alpha_{\ell m}B_n(t)\mathcal{Y
 }_{n\ell m}(\chi,\theta,\phi)
\end{align*}
The plasma velocity can be expressed in terms of the velocity potential and then expand as
\begin{align*}
 \delta u_k(\textbf{x},t)&=\nabla_k\,\delta u(\textbf{x},t)\\
 &=\sum_{n\ell m}\alpha_{\ell m}\delta u_n(t)\nabla_k\,\mathcal{Y}_{n\ell m}(\chi,\theta,\phi)
\end{align*}
Following the technique used in the flat case\cite{r24}, There is no preferred direction in the problem except initial conditions which are encapsulated in the stochastic parameter $\alpha_{\ell m}$. So the coefficient of $\alpha_{\ell m}$ in the intensity matrix can be decomposed into a sum of terms proportional to the two symmetric tensors $\left(\tilde{g}^{ij}-\hat{p}^i\hat{p}^j\right)$ and $(\nabla^i-\hat{p}^i\hat{p}_s\nabla^s)(\nabla^j-\hat{p}^j\hat{p}_t\nabla^t)$ that vanishes when contracted with $\hat{p}^i$ or $\hat{p}^j$, with coefficients $\Delta_{Tn}$ and $\Delta_{Pn}$ that depend on the directions $\hat{p}$ and $\hat{q}=k_n^{-1}\nabla$ only through the scalar product $\hat{q}\cdot\hat{p}$. So we may expand $J^{ij} (\textbf{x},\hat{p},t)$  matrix as
\begin{align}\label{274}
  J^{ij} (\textbf{x},\hat{p},t)=&\sum_{n\ell m}\alpha_{\ell m}\bigg[\frac{1}{2}\left(\Delta_{Tn}(\hat{q}\cdot\hat{p},t)-\Delta_{Pn}(\hat{q}\cdot\hat{p},t)\right)\times\nonumber\\
  &\left(\tilde{g}^{ij}-\hat{p}^i\hat{p}^j\right)\mathcal{Y}_{n\ell m}(\chi,\theta,\phi)\nonumber\\
  +&\,\Delta_{Pn}(\hat{q}\cdot\hat{p},t)k_n^{-2}q^{ij}\mathcal{Y}_{n\ell m}(\chi,\theta,\phi)\bigg]
\end{align}
where
\begin{equation*}
  q^{ij}=\frac{(\nabla^i-\hat{p}^i\hat{p}_s\nabla^s)(\nabla^j-\hat{p}^j\hat{p}_t\nabla^t)}{\mathcal{Y}^{-1}_{n\ell m}(\chi,\theta,\phi)k_n^{-2}(\nabla^2-\hat{p}_s\hat{p}_tH^{st})\mathcal{Y}_{n\ell m}(\chi,\theta,\phi)}
\end{equation*}
Note that the trace $J^i_{\;i}(\textbf{x},\hat{p},t)$, which we will derive the temperature fluctuation from that, equals to
\begin{equation*}
  J^i_{\;i}(\textbf{x},\hat{p},t)=\sum_{n\ell m}\alpha_{\ell m}\Delta_{Tn}(\hat{q}\cdot\hat{p},t)\mathcal{Y}_{n\ell m}(\chi,\theta,\phi)
\end{equation*}
Also, we introduce the source functions $\varphi_n (t)$ and $\mathcal{J}_n (t)$ as
\begin{align}\label{275}
 \int\frac{\dd[2]{\hat{p}_1}}{4\pi}&\sqrt{Det\,\tilde{g}^{ij}}J^{ij}(x,p_1,t)\\
 &=\sum_{n\ell m}\alpha_{\ell m}\Big[\varphi_n(t)\tilde{g}^{ij}\mathcal{Y}_{n\ell m}(\chi,\theta,\phi)\\
 &-\frac{1}{2}\mathcal{J}_n(t)H^{ij}k_n^{-2}\mathcal{Y}_{n\ell m}(\chi,\theta,\phi)\Big]
\end{align}

In this particular coordinate $(\chi,\theta,\phi)$, the momentum $\hat{p}$ for the photon coming from direction $\hat{n}$ will be $ \hat{p}=-\hat{n}=(-1,0 ,0)=-e_\chi$.\\
Because of the conditions $\hat{p}_i\,J^{ij}(\textbf{x},\hat{p},t)=0$ and\\ $\hat{p}_j\,J^{ij}(\textbf{x},\hat{p},t)=0$, this is just the $\chi\chi$ element of the $J^{ij}(\textbf{x},\hat{p},t)$ matrix that contributes to the fluctuation calculations. So, for $\hat{p}^s\hat{p}^tH_{st},\; \hat{p}^k\nabla_k$ and\\ $\hat{p}^k\hat{p}^m\hat{p}^n\partial_k\tilde{g}_{mn}$ we can write
\begin{align*}
 &\hat{p}^k\nabla_k=-\nabla_\chi\qquad\hat{p}^s\hat{p}^tH_{st}=\nabla_\chi\nabla_\chi\\
 &\hat{p}^k\hat{p}^m\hat{p}^n\partial_k\tilde{g}_{mn}=-\partial_\chi \tilde{g}_{\chi\chi}=0
\end{align*}
Inserting all above relations in Eq.(\ref{260}), one can show that the Boltzmann equation for the matrix\\
$J^{ij}(\textbf{x},\hat{p},t)$ yields two coupled Boltzmann equations for $\Delta_{Tn}(\mu,t)$ and $\Delta_{Pn}(\mu,t)$ as \cite{r28}
\begin{align}\label{28}
  &\dot{\Delta}_{Tn}(\mu,t)\Pi_{n \ell}(\chi)-\Delta_{Tn}(\mu,t)\frac{1}{a(t)}\dv{\chi}\Pi_{n \ell}(\chi)\nonumber\\
  &+2\dot{A}_n(t)\Pi_{n \ell}(\chi)+2\dot{B}_n(t)\dv[2]{\chi}\Pi_{n \ell}(\chi)\nonumber\\
  &=-\omega_c(t)\Delta_{Tn}(\mu,t)\Pi_{n\ell}(\chi)+3\,\omega_c(t)\varphi_n(t)\Pi_{n\ell}(\chi)\nonumber\\
  &-\frac{4\,\omega_c(t)}{a(t)}\delta u_n(t)\dv{\chi}\Pi_{n\ell}(\chi)\nonumber\\
  &+\frac{3}{4}\omega_c(t)\mathcal{J}_n(t)\left(1+k_n^{-2}\dv[2]{\chi}\right)\Pi_{n\ell}(\chi)
\end{align}

\begin{align}\label{281}
  &\dot{\Delta}_{Pn}(\mu,t)\Pi_{n \ell}(\chi)-\Delta_{Pn}(\mu,t)\frac{1}{a(t)}\dv{\chi}\Pi_{n \ell}(\chi)\nonumber\\
  &=-\omega_c(t)\Delta_{Pn}(\mu,t)\Pi_{n\ell}(\chi)\nonumber\\
  &+\frac{3}{4}\omega_c(t)\mathcal{J}_n(t)\left(1+k_n^{-2}\dv[2]{\chi}\right)\Pi_{n\ell}(\chi)
\end{align}
where again $\Pi_{n\ell} (\chi)$ is the hyperspherical Bessel function that has been introduced in section \ref{602}.\\
For a photon coming from direction $\hat{n}$ we have $\dd t=-a(t)\dd \chi$, so
\begin{align*}
  \dv{\chi}&=-a(t)\dv{t}\nonumber\\
  \dv[2]{\chi}&=-a(t)\dot{a}(t)\dv{t}+a^2(t)\dv[2]{t}\nonumber\\
  \chi(t)&=\int_{t}^{t_0}\frac{\dd t'}{a(t')}
\end{align*}
By using above relations and also integrating by parts of Eqs.(\ref{28}),(\ref{281}), one can get the line of sight solution for $\Delta_{Tn}(\mu,t)$ as
\begin{subequations}\label{30}
  \begin{equation}\label{30-a}
  \Delta_{Tn}(t)\Pi_{n\ell} (\chi)=\int_{t_1}^{t}\exp\left(-\int_{t'}^{t}\dd t''\omega_c(t'')\right)\mathcal{G}_{n \ell}(t')\dd t'
\end{equation}
\begin{align}\label{30-b}
 \mathcal{G}_{n\ell}=&\left(-2\dot{A}_n+3\,\omega_c(t)\varphi_n(t)+\frac{3}{4}\omega_c(t)\mathcal{J}_n(t)\right)\Pi_{n\ell}(t)\nonumber\\
 +&\bigg(4\,\omega_c(t)\delta u_n(t)-2a\dot{a}\dot{B}_n\nonumber\\
 +&\frac{3}{4}\omega_c(t)\,\mathcal{J}_n(t)\,a\dot{a}k_n^{-2}\bigg)\dv{t}\Pi_{n\ell}(t)\nonumber\\
 +&\left(-2a^2\dot{B}_n+\frac{3}{4}\omega_c(t)\mathcal{J}_n(t)a^2k_n^{-2}\right)\dv[2]{t}\Pi_{n\ell}(t)
\end{align}
\end{subequations}
where $t_1$ is a time that we choose it to be sufficiently early, so that $\omega_c (t_1 )$ is much bigger than the expansion rate of the universe and $t$ at any time after recombination.
Also the line of sight solution for $\Delta_{Pn}(t)$ would be
\begin{subequations}\label{310}
  \begin{equation}\label{310-a}
  \Delta_{Pn}(t)\Pi_{n\ell} (\chi)=\int_{t_1}^{t}\exp\left(-\int_{t'}^{t}\dd t''\omega_c(t'')\right)\mathcal{H}_{n \ell}(t')\dd t'
\end{equation}
\begin{align}\label{310-b}
 \mathcal{H}_{n\ell}=&\frac{3}{4}\omega_c(t)\mathcal{J}_n(t)\bigg(\Pi_{n\ell}(t)+k_n^{-2}a\dot{a}\dv{t}\Pi_{n\ell}(t)\nonumber\\
 +&k_n^{-2}a^2\dv[2]{t}\Pi_{n\ell}(t)\bigg)
\end{align}
\end{subequations}
The temperature fluctuation at our position $\textbf{x}=0$ and time $t=t_0$ can be written as
\begin{align}\label{34}
\Big(\Delta T(\hat{n})\Big)^S=&\frac{T_0}{4}J^i_{\;i}(\textbf{x}=0,\hat{p},t_0)\nonumber\\
=&\frac{T_0}{4}\sum_{n\ell m}\alpha_{\ell m}\Delta_{Tn}(t_0)\mathcal{Y}_{n\ell m}(\chi=0,\theta,\phi)\nonumber\\
=&\frac{T_0}{4}\sum_{n\ell m}\alpha_{\ell m}\Delta_{Tn}(t_0)\Pi(t_0)Y_{\ell m}(\theta,\phi)\nonumber\\
=&\sum_{\ell m}a^S_{T,\ell m}Y_{\ell m}(\theta,\phi)
\end{align}
where the scalar contribution to the multipole coefficient  $a_{T,\ell m}^S$ is
\begin{equation}\label{35}
  a^S_{T,\ell m}=\frac{T_0}{4}\sum_{n}\alpha_{\ell m}\int_{t_1}^{t_0}\exp\left(-\int_{t}^{t_0}\dd t'\omega_c(t')\right)\mathcal{G}_{n\ell}(t)\dd t
\end{equation}
By Assuming sudden transition from opacity to transparency at the time of last scattering, considering the evolution of perturbation hydrodynamically and also neglecting the $ISW$ effect that is important only for relatively small values of $\ell$, one can derive the final formula for the scalar mode temperature fluctuations as \cite{r28}
\begin{subequations}\label{39}
\begin{align}\label{39a}
  \Big(\Delta T(\hat{n})\Big)^S_{early}=\sum_{\ell m}a^S_{T,\ell m}Y_{\ell m}(\theta,\phi)
\end{align}
where
\begin{align}\label{39b}
  a^S_{T,\ell m}=T_0\,\sum_{n}\alpha_{\ell m}\bigg[F_n\Pi_{n\ell}(\chi_L)+G_n\dv{\chi_L}\Pi_{n\ell}\bigg]
\end{align}
\end{subequations}
and
\begin{align}\label{70}
  F_n=&-\frac{1}{2}a^2(t_L)\ddot{B}(t_L)-\frac{1}{2}a(t_L)\dot{a}(t_L)\dot{B}_n(t_L)+\frac{\delta T_n(t_L)}{\overline{T}(t_L)}\nonumber\\
  =&\frac{1}{3}\delta_\gamma(t_L)-\frac{a^2(t_L)}{3t_L\,q_n^2}\psi_n(t_L)\nonumber\\
  =&\frac{\mathcal{R}_n^o}{5}\Bigg[3\mathcal{T}(\kappa_n)R_L\nonumber\\
  -&(1+R_L)^{-1/4}e^{-\int_{0}^{t_L}\Gamma\dd t}\mathcal{S}(\kappa_n)\times\nonumber\\
  &\cos\bigg(\int_{0}^{t_L}\frac{q_n\,\dd t}{a\sqrt{3(1+R)}}+\Delta(\kappa_n)\bigg)\Bigg]
\end{align}
\begin{align}\label{71}
  G_n=&-\left(\frac{1}{2}a(t_L)\dot{B}(t_L)+\frac{\delta u_n(t_L)}{a(t_L)}\right)\nonumber\\
  =&-\left(-\frac{a(t_L)}{q_n^2}\psi_n(t_L)+\frac{\delta u_n(t_L)}{a(t_L)}\right)\nonumber\\
  =&-\frac{\sqrt{3}\mathcal{R}_n^o}{5q_n(1+R_L)^{3/4}} e^{-\int_{0}^{t_L}\Gamma\dd t}\mathcal{S}(\kappa_n)\times\nonumber\\
  &\sin\bigg(\int_{0}^{t_L}\frac{q_n\,\dd t}{a\sqrt{3(1+R)}}+\Delta(\kappa_n)\bigg)
\end{align}
where $R=\frac{3\,\overline{\rho}_B}{4\,\overline{\rho}_\gamma}$, $\mathcal{R}_n^o$ is the fourier component of comoving curvature perturbation (the superscript $o$ standing for "outside the Hubble horizon") and $\mathcal{T}(\kappa_n ),\; \mathcal{S}(\kappa_n ),\; \Delta(\kappa_n )$ are time-independent dimensionless function of the dimensionless re-scaled wave number $\kappa_n=\frac{\sqrt{2}\,q_n}{a_{EQ}H_{EQ}}$, called transfer functions. A detail expression for these functions and also acoustic damping rate $\Gamma(t)$ could be found in \cite{r24}.\\
%%%%%%%%%%%%%%%%%%%%%%%%%%%%%%%%%%%%%%%%%%%%%%%%%%%%%%%%%%%%%%%%%%%%%%%%%%%%%%%%%%%%%%%%%%%%%%%%%%%%%%%%%%%%%%%%%%%%%%%%%%%%%%%%%%%%%%%%%%%%%%%%%%%%%%%%%%
%Section 3
\section{CMB Polarization from the Boltzmann equation formalism}\label{603}
In the previous section, we extracted the scalar mode temperature multipole moment $a^S_{T,\ell m}$. Now, we are going to extract the multipole moment of the CMB polarization in a spatially closed background which will be used to compute the EE and TE multipole coefficients.\\
We define the Stokes parameters for a photon coming from an arbitrary direction $\hat{n}$ as
\begin{subequations}\label{13}
  \begin{equation}\label{13-a}
  Q^S(\hat{n})\pm \, iU^S(\hat{n})=\frac{T_0}{2}e_{\pm i}(\hat{n})e_{\pm j}(\hat{n})J^{ij}(0,-\hat{n},t_0)
  \end{equation}
  \begin{align}\label{13-b}
  V^S(\hat{n})=&\frac{T_0}{4}e_{-i}(\hat{n})e_{+j}(\hat{n})\times\nonumber\\
  &\bigg(J^{ij}(0,-\hat{n},t_0)-J^{ji}(0,-\hat{n},t_0)\bigg)
  \end{align}
\end{subequations}
where $e_{\pm}(\hat{n})$ are the polarization vectors for a photon coming from the direction $\hat{n}$.
The scattering of light by non-relativistic electrons does not produce circular polarization, and therefore we expect that all microwave background photons are linearly polarized, so that $J^{ij}$ is real and $V=0$.\\
Using the expansion of $J^{ij}$ introduced in the previous section, we can write
\begin{align}\label{14}
  &Q^S(\hat{n})\pm \, iU^S(\hat{n})=\frac{T_0}{2}e_{\pm i}(\hat{n})e_{\pm j}(\hat{n})J^{ij}(0,-\hat{n},t_0)\nonumber\\
  &=\frac{T_0}{2}\sum_{n\ell m}\alpha_{\ell m}\bigg[\frac{1}{2}\big(\Delta_{Tn}(t)-\Delta_{Pn}(t)\big)\times\nonumber\\
  &e_{\pm i}(\hat{n})e_{\pm j}(\hat{n})\left(\tilde{g}^{ij}-\hat{n}^i\hat{n}^j\right)\mathcal{Y}_{n\ell m}(\chi=0,\theta,\phi)\nonumber\\
  &+\,\Delta_{Pn}(t_0)e_{\pm i}(\hat{n})e_{\pm j}(\hat{n})k_n^{-2}q^{ij}\mathcal{Y}_{n\ell m}(\chi=0,\theta,\phi)\bigg]
\end{align}
Since $\hat{n}^ie_{\pm i}=0$ and $e_{\pm i}e^{\pm i}=0$, the only term that contributes to the Stokes parameters is the one proportional to $q^{ij}$:
\begin{align}\label{15}
  &Q^S(\hat{n})\pm \, iU^S(\hat{n})=\nonumber\\
  &\frac{T_0}{2}\sum_{n\ell m}\alpha_{\ell m}\Delta_{Pn}(t_0)e_{\pm i}(\hat{n})e_{\pm j}(\hat{n})k_n^{-2}q^{ij}\mathcal{Y}_{n\ell m}(\chi=0,\theta,\phi)\nonumber\\
  &=\frac{T_0}{2}\sum_{n\ell m}\alpha_{\ell m}e_{\pm i}(\hat{n})e_{\pm j}(\hat{n})\times\nonumber\\
  &\frac{\nabla^i\nabla^j}{\mathcal{Y}^{-1}_{n\ell m}k_n^{-2}(\nabla^2-\hat{n}_m\hat{n}_nH^{mn})\mathcal{Y}_{n\ell m}}\times\nonumber\\
  &\,k_n^{-2}\Delta_{Pn}(t_0)\Pi_{n\ell}(\chi=0)\,Y_{\ell m}(\theta,\phi)
\end{align}

By using Eq.(\ref{310}), the above relation then reads
\begin{align}\label{17}
  &Q^S(\hat{n})\pm \, iU^S(\hat{n})=\nonumber\\
  &\frac{T_0}{2}\sum_{n\ell m}\alpha_{\ell m}e_{\pm i}(\hat{n})e_{\pm j}(\hat{n})\times\nonumber\\
  &\frac{\nabla^i\nabla^j}{\mathcal{Y}^{-1}_{n\ell m}k_n^{-2}(\nabla^2-\hat{n}_m\hat{n}_nH^{mn})\mathcal{Y}_{n\ell m}}\times\nonumber\\
  &\,k_n^{-2}\int_{t_1}^{t_0}\exp\left(-\int_{t}^{t_0}\dd t'\omega_c(t')\right)\mathcal{H}_{n \ell}(t)\dd t\,Y_{\ell m}(\theta,\phi)\nonumber\\
  =&\frac{-3T_0}{8}\sum_{n\ell m}\alpha_{\ell m}e_{\pm i}(\hat{n})e_{\pm j}(\hat{n})\times\nonumber\\
  &k_n^{-2}\nabla^i\nabla^j\int_{t_1}^{t_0}P(t)\,\mathcal{J}_n(t)\,\Pi_{n\ell}(t)\,\dd t\,Y_{\ell m}(\theta,\phi)
\end{align}
where
\begin{equation*}
P(t)=\omega_c(t)\exp\left(-\int_{t}^{t_0}\dd t'\,\omega_c(t')\right)
\end{equation*}
We introduce the angular gradient operator acting on $\hat{n}$ as
\begin{align*}
  \nabla^i&=\hat{n}^i\pdv{}{r}+\frac{\widetilde{\nabla}^i}{r}\qquad\widetilde{\nabla}=\hat{\theta}\pdv{}{\theta}+\frac{\hat{\phi}}{\sin\theta}\pdv{}{\phi}
\end{align*}
By using the relations $\hat{n}^ie_{\pm i}=0$ and $e_{\pm i}e^{\pm i}=0$ we can replace the gradient operator by angular one as
\begin{align*}
  e_{\pm i}(\hat{n})e_{\pm j}(\hat{n})\nabla^i\,\nabla^j&=e_{\pm i}(\hat{n})e_{\pm j}(\hat{n})\times\nonumber\\
  &(\hat{n}^i\pdv{}{r}+\frac{\widetilde{\nabla}^i}{r})(\hat{n}^j\pdv{}{r}+\frac{\widetilde{\nabla}^j}{r})\\
  &=e_{\pm i}(\hat{n})e_{\pm j}(\hat{n})\frac{\tilde{\nabla}^i\,\tilde{\nabla}^j}{r^2}
\end{align*}
Now, by introducing the Spin-2 harmonic oscillators as
\begin{align}\label{18}
  {}_{2}Y_{\ell m}(\theta,\phi)=2\sqrt{\frac{(\ell-2)!}{(\ell+2)!}}\,e_{+i}(\hat{n})e_{+j}(\hat{n})\tilde{\nabla}^i\,\tilde{\nabla}^j\,Y_{\ell m}(\theta,\phi)
\end{align}
we can write Eq.(\ref{17}) as
\begin{align}\label{19}
  &Q^S(\hat{n})+ \, iU^S(\hat{n})=-\frac{3T_0}{16}\sum_{n\ell m}\alpha_{\ell m}\sqrt{\frac{(\ell-2)!}{(\ell+2)!}}\times\nonumber\\
  &\int_{t_1}^{t_0}P(t)\,\mathcal{J}_n(t)\,\frac{\Pi_{n\ell}(t)}{k_n^2\,r^2}\,\dd t\,{}_{2}Y_{\ell m}(\theta,\phi)
\end{align}

On the other hand, we should expand the Stokes parameters in terms of Spin-2 harmonic oscillators because of the same dependence on the polarization vectors $e_{\pm}(\hat{n})$ that gives the same behavior under rotations \cite{r24}. So for $Q^S(\hat{n})+ \, iU^S(\hat{n})$ we can write
\begin{align}\label{20}
  Q^S(\hat{n})+ \, iU^S(\hat{n})=\sum_{\ell m}a^S_{P,\ell m}\,{}_{2}Y_{\ell m}(\theta,\phi)
\end{align}
where the multipole moment of polarization $a^S_{P,\ell m}$ equals to
\begin{align}\label{21}
 a^S_{P,\ell m}=&-\frac{3T_0}{16}\sqrt{\frac{(\ell-2)!}{(\ell+2)!}}\times\nonumber\\
 &\sum_{n}\alpha_{\ell m}\int_{t_1}^{t_0}\dd t\, P(t)\mathcal{J}_n(t)\,\frac{\Pi_{n\ell}(t)}{k_n^2\,r^2}
\end{align}

By using Eqs.(\ref{274}) and (\ref{275}) that give definitions for $J^{ij}$, $\varphi_n (t)$ and $\mathcal{J}_n (t)$, one can find
\begin{align}\label{23}
 \alpha_{\ell m}^\ast\mathcal{J}_n(t)^\ast=\alpha_{\ell -m}\,\mathcal{J}_n(t)
\end{align}
and therefore
\begin{align}\label{24}
a^{S\ast}_{P,\ell m}=a^S_{P,\ell -m}
\end{align}
It is conventional to define coefficients $a^S_{E,\ell m}$ and $a^S_{B,\ell m}$ because of their useful properties under space inversion as\cite{r24}
\begin{align}\label{25}
  a^S_{E,\ell m}&=-(a^S_{P,\ell m}+a^{S\ast}_{P,\ell m})/2\\
  a^S_{B,\ell m}&=i\,(a^S_{P,\ell m}-a^{S\ast}_{P,\ell m})/2
\end{align}
Inspection of above equations and the relation $a^{S\ast}_{P,\ell m}=a^S_{P,\ell -m}$ then shows that the scalar modes contributes only an E-type polarization
\begin{align}\label{26}
  a^S_{E,\ell m}&=-a^S_{P,\ell m}\\
  a^S_{B,\ell m}&=i\,(a^S_{P,\ell m}-a^{S\ast}_{P,\ell m})/2=0
\end{align}
and $a^S_{E,\ell m}$ equals to
\begin{align}\label{27}
  &a^S_{E,\ell m}=-a^S_{P,\ell m}\nonumber\\
  &=\frac{3\,T_0}{16}\sqrt{\frac{(\ell+2)!}{(\ell-2)!}}\,\sum_{n}\alpha_{\ell m}\int_{t_1}^{t_0}\dd t\, P(t)\,\mathcal{J}_n(t)\frac{\Pi_{n\ell}(t)}{k_n^2\,r^2}
\end{align}
By assuming that $\omega_c (t)$ drops sharply at time $t_L$ from a value much greater than the expansion rate to zero (a sudden transition from opacity to transparency), the integral $\int_{t_1}^{t_0}\dd t\,\omega_c(t)\exp\left(-\int_{t}^{t_0}\dd t'\,\omega_c(t')\right)$ is non zero only in a narrow interval around  $t_L$. Also under the same assumption, the factor $\exp\left(-\int_{t}^{t_0}\dd t'\,\omega_c(t')\right)$ rise sharply from zero for time before recombination to unity for $t>t_L$.
Using above approximation $a^S_{E,\ell m}$ can be written as
\begin{align}\label{270}
  &a^S_{E,\ell m}=\nonumber\\
  &\frac{3\,T_0}{16}\sqrt{\frac{(\ell+2)!}{(\ell-2)!}}\,\sum_{n}\alpha_{\ell m}\,\mathcal{J}_n(t_L)\frac{\Pi_{n\ell}(t_L)}{k_n^2\,r_L^2}
\end{align}

%%%%%%%%%%%%%%%%%%%%%%%%%%%%%%%%%%%%%%%%%%%%%%%%%%%%%%%%%%%%%%%%%%%%%%%%%%%%%
%Section 4
\section{The analytical EE and TE multipole coefficients}\label{604}
We can now give formulas for the scalar contribution to the EE and TE multipole coefficients which are defined as
\begin{align}\label{100}
  &\langle a^{S\ast}_{E,\ell m} \, a^S_{E,\ell' m'}\rangle=C_{EE}\delta_{\ell \ell'}\delta_{m m'}\\
  &\langle a^{S\ast}_{T,\ell m} \, a^S_{E,\ell' m'}\rangle=C_{TE}\delta_{\ell \ell'}\delta_{m m'}
\end{align}
Recalling the normalization condition $ \langle\alpha_{\ell m}\,\alpha_{\ell' m'}^\ast\rangle=\delta_{\ell\ell'}\,\delta_{mm'}$ and $\langle A_n(t)\,A_{n'}^\ast(t)\rangle=A_n^2(t)\,\delta_{nn'}$ we see from Eq.(\ref{100}) that
\begin{align}\label{29}
  C_{EE}&=\frac{9\,T_0^2}{16^2}\frac{(\ell+2)!}{(\ell-2)!}\,\sum_{n}\bigg[\mathcal{J}_n(t_L)\frac{\Pi_{n\ell}(\chi_L)}{k_n^2\,r_L^2}\bigg]^2
\end{align}
Now, we should find an expression for $\mathcal{J}_n(t_L)$ in terms of known perturbations. By using the definition of $J^{ij}$, $\varphi_n (t)$ and $\mathcal{J}_n (t)$ one can write $\mathcal{J}_n (t)$ as
\begin{align}\label{30}
 \mathcal{J}_n(t)&=\frac{1}{2}\int\frac{\dd[2]{\hat{p}}}{4\pi}\sqrt{Det\,\tilde{g}^{ij}}\Delta_{Tn}\bigg(1+3k_n^{-2}(\hat{p}_m\nabla^m)^2\bigg)\nonumber\\
 &+\frac{3}{2}\int\frac{\dd[2]{\hat{p}}}{4\pi}\sqrt{Det\,\tilde{g}^{ij}}\Delta_{Pn}\bigg(1+k_n^{-2}(\hat{p}_m\nabla^m)^2\bigg)
\end{align}
in which the term including $\Delta_{Pn}$ take account of whatever polarization the photon may already have at the time of last scattering and can be expected to be relatively small since, under the condition of rapid photon scattering, photons must be unpolarized.\\
On the other hand, $\delta T^i_{\gamma j}$ can be written as
\begin{align}\label{31}
 &\delta T^i_{\gamma j}(x,t)=\bar{\rho}_\gamma (t)\int\frac{\dd[2]{\hat{p}}}{4\pi}\sqrt{Det\,\tilde{g}^{ij}}J^m_m(x,\hat{p},t)\hat{p}^i\hat{p}_j\nonumber\\
 &=\frac{1}{2}\tilde{g}^i_{\,j}\,\bar{\rho}_\gamma (t)\int\frac{\dd[2]{\hat{p}}}{4\pi}\sqrt{Det\,\tilde{g}^{ij}}\times\nonumber\\
 &\sum_{n\ell m}\alpha_{\ell m}\Delta_{Tn}\bigg(1+k_n^{-2}(\hat{p}_m\nabla^m)^2\bigg)\mathcal{Y}_{n\ell m}\nonumber\\
 &+\frac{1}{2}k_n^{-2}\nabla^i\nabla_j\,\bar{\rho}_\gamma(t)\int\frac{\dd[2]{\hat{p}}}{4\pi}\sqrt{Det\,\tilde{g}^{ij}}\times\nonumber\\
 &\sum_{n\ell m}\alpha_{\ell m}\Delta_{Tn}\bigg(1+3k_n^{-2}\,(\hat{p}_m\nabla^m)^2\bigg)\mathcal{Y}_{n\ell m}\nonumber\\
 &=\frac{1}{2}\tilde{g}^i_{\,j}\,\bar{\rho}_\gamma (t)\int\frac{\dd[2]{\hat{p}}}{4\pi}\sqrt{Det\,\tilde{g}^{ij}}\times\nonumber\\
 &\sum_{n\ell m}\alpha_{\ell m}\Delta_{Tn}\bigg(1+k_n^{-2}\,(\hat{p}_m\nabla^m)^2\bigg)\mathcal{Y}_{n\ell m}\nonumber\\
 &+k_n^{-2}\nabla^i\nabla_j\,\bar{\rho}_\gamma \sum_{n\ell m}\alpha_{\ell m}\,\mathcal{J}_n(t)\mathcal{Y}_{n\ell m}
\end{align}
By comparing this equation with the relation $\delta T^i_{\gamma j}(x,t)=\tilde{g}^i_{\,j} \delta P + \nabla^i\nabla_j\,\pi^S $ \cite{r24} we have
\begin{align}\label{32}
\mathcal{J}_n(t)=\frac{k_n^2\,\pi^S_n(t) }{\bar{\rho}_\gamma}
\end{align}
For the anisotropy inertia $\pi^S_n$ we can write \cite{r24}
\begin{equation}\label{33}
  \nabla^i\nabla^j\pi_n^S(t)=-a^{-2}\eta_\gamma \nabla^i\nabla^j \,\delta \tilde{u}_n(t)
\end{equation}
where $\eta_\gamma$ is the shear viscosity due to photon momentum transport, given in terms of photon mean free time $t_\gamma$ by $\eta_\gamma=\frac{16}{45}\bar{\rho}_\gamma t_\gamma$, and $\delta \tilde{u}$ is the gauge invariant velocity potential
\begin{align}\label{34}
  \delta \tilde{u}_n=\delta u_n-a\,F_n+\frac{a^2\dot{B}_n}{2}
\end{align}
with $F_n$ and $B_n$ metric perturbations. So in synchronous gauge (where $F_n=0$) we have
\begin{align}\label{35}
  \pi^S_n(t) =-a^{-1}(t)\,\frac{16}{45}\,\bar{\rho}_\gamma\, t_\gamma\,G_n(t)
\end{align}
and finally conclude that
\begin{align}\label{36}
  \mathcal{J}_n(t_L)=\frac{16\, k_n^2}{45 a}\,\bar{t}_\gamma\, G_n(t_L)
\end{align}
where as before
\begin{align}\label{37}
  G_n(t_L)=-\left(\frac{1}{2}a(t_L)\dot{B}(t_L)+\frac{\delta u_n(t_L)}{a(t_L)}\right)
\end{align}
and $\bar{t}_\gamma$ is some appropriate average of $t_\gamma$ during the end of recombination era (From $Z\simeq1000$ till $Z\simeq1100$).
By using above relations, we can write $C^S_{EE,\ell}$ as
\begin{align}\label{39}
  C^S_{EE,\ell}=\frac{9T_0^2}{45^2}\frac{(\ell+2)!}{(\ell-2)!}\frac{\bar{t}_\gamma^2}{a^2_L r^4_L}\sum_{n}G^2_n(t_L)\Pi_{n\ell}^2(\chi_L)
\end{align}
and as a reminder
\begin{align*}
  G_n(t_L)=&-\frac{\sqrt{3}\mathcal{R}_n^o}{5q_n(1+R_L)^{3/4}} e^{-\int_{0}^{t_L}\Gamma\dd t}\mathcal{S}(\kappa_n)\times\nonumber\\
  &\sin\bigg(\int_{0}^{t_L}\frac{q_n\,\dd t}{a\sqrt{3(1+R)}}+\Delta(\kappa_n)\bigg)
\end{align*}

In order to recover the sudden transition from opacity to transparency assumption, we multiply $\sin$ function (and also $\cos$ function for TE multipole calculation) appeared in the form factor with a Gaussian probability function and average it in time \cite{r24}. The whole effect of this averaging is to introduce an additional damping factor $\exp(-\omega_L^2\sigma_t^2/2)$ which can be added to the acoustic damping factor
\begin{align}\label{41}
 \int_{0}^{t_L}\Gamma \dd t+\frac{\omega_L^2\sigma_t^2}{2}=\left(\frac{q_nd_D}{a_L}\right)^2
\end{align}
and introduce a new parameter $d_D$ called “damping length” as
\begin{subequations}\label{42}
  \begin{align}\label{42-a}
  d_D^2=d_{Silk}^2+d_{Landau}^2
\end{align}
where
\begin{align}\label{42-b}
  d_{Silk}^2=a^2_L\int_{0}^{t_L}\frac{t_\gamma\, c^2\, \dd t}{6a^2(1+R)}\left[\frac{16}{15}+\frac{R^2}{1+R}\right]
\end{align}
\begin{align}\label{42-c}
  d_{Landau}^2=\frac{\sigma_t^2}{6(1+R_L)}
\end{align}
\end{subequations}
Here, $R=3\bar{\rho}_B/4\bar{\rho}_\gamma$ and $R_L=3\Omega_B/4\Omega_\gamma(1+z_L)$, $\sigma_t$ is the standard deviation in the time of last scattering which is related to the standard deviation $\sigma$ in the temperature of last scattering by $\sigma_t=3\,t_L\sigma/2\,T_L$ and $t_\gamma$ is the photon mean free time which can be written as
\begin{align*}
  t_\gamma=\frac{1}{\omega_c}=\frac{R^3}{n_{B0}R_0^3(1-Y)X\sigma_{\mathcal{T}}c}
\end{align*}
where $R_0=3\Omega_B/4\Omega_\gamma$ is the present value of $R$, $Y\simeq0.24$ is the fraction of nucleons in the form of un-ionized helium around the time of last scattering, $n_{B0}=3H_0^2\Omega_B/8\pi Gm_N$ is the present number density of baryons, $\sigma_{\mathcal{T}}$ is the cross section for Thomson scattering and $X(R)$ is the fractional ionization, calculated in Ref. \cite{r24}. Also,
\begin{align}\label{430}
  \dd t&=\frac{\dd R}{R\,H_0\sqrt{\Omega_M(R_0/R)^3+\Omega_R(R_0/R)^4}}\nonumber\\
  &=\frac{R\,\dd R}{H_0\sqrt{\Omega_M}R_0^{3/2}\sqrt{R_{EQ}+R}}
\end{align}
where $R_{EQ}=(\Omega_R/\Omega_M)R_0=3\Omega_R\Omega_B/4\Omega_M\Omega\gamma$ is the value of $R$ at matter-radiation equality. Putting all above relations together, the Silk and Landau damping length are given by
\begin{align}\label{431}
  d_{Silk}^2&=\frac{R_L^2\,c^2}{6\,(1-Y)\,n_{B0}\sigma_{\mathcal{T}}\,c\,H_0\sqrt{\Omega_M}\,R_0^{9/2}}\times\nonumber\\
  &\int_{0}^{t_L}\frac{R^2\dd R}{X(R)(1+R)\sqrt{R_{EQ}+R}}\left[\frac{16}{15}+\frac{R^2}{1+R}\right]
\end{align}
and
\begin{align}\label{432}
  d_{Landau}^2=\frac{3\,\sigma^2\,t_L^2}{8\,T_L^2\,(1+R_L)}
\end{align}
We also introduce $d_H$ and $d_T$ as
\begin{align}\label{43}
  d_H&=a_L\int_{0}^{t_L}\frac{c\, \dd t}{a\sqrt{3(1+R)}}\nonumber\\
  &=\frac{2}{H_0\sqrt{3R_L\Omega_M}(1+z_L)^{3/2}}\times\nonumber\\
  &\ln\left(\frac{\sqrt{1+R_L}+\sqrt{R_{EQ}+R_L}}{1+\sqrt{R_{EQ}}}\right)
\end{align}
and
\begin{align}\label{44}
  d_T=\frac{0.0177}{\Omega_M h^2}\qquad \kappa_n=\frac{q_nd_T}{a_L}
\end{align}
Also, from the theory of inflation, we know $\mathcal{R}^o_n\propto q_n^{-3/2}$ \cite{r26}. Following the parameterizations we have used for this quantity in $C^S_{TT,\ell}$ derivation\cite{r28}, we can write
\begin{align}\label{45}
  |\mathcal{R}^o_n|^2=N^2\,q_n^{-3}\left(\frac{q_n/a_0}{k_\mathcal{R}}\right)^{n_s-1}
\end{align}
By putting all above relations into Eq.(\ref{39}) and using the WKB approximation for $\Pi_{n\ell}(\chi_L)$ \cite{r28}, one arrives at the explicit, analytical formula for the scalar contribution to the EE multipole coefficient as
\begin{align}\label{46}
&\frac{\ell(\ell+1)}{2\pi}C^S_{EE,\ell}=\frac{4\pi\,T_0^2\,N^2}{25^2\times3}\frac{\bar{t}_\gamma^2}{r^2_L}(1+z_L)^2\times\nonumber\\
&\sum_{n}\frac{\ell^5\left(\frac{n/a_0}{\alpha\, k_R}\right)^{n_s-1}}{n^4\sin^4\chi_L\big[\frac{n^2}{\ell^2}-\frac{1}{\sin\chi_L^2}\big]^{1/2}(1+R_L)^{3/2}}\nonumber\\
&e^{-2\left(\frac{q_nd_D}{a_L}\right)^2}S^2\left(\frac{q_nd_T}{a_L}\right)\sin^2\left(\frac{q_nd_H}{a_L}+\Delta\left(\frac{q_nd_T}{a_L}\right)\right)
\end{align}
where
\begin{subequations}\label{50}
\begin{align}\label{50a}
 \sin\chi_L=\sin\bigg[\frac{c}{\alpha\, a_0\,H_0}\int_{\frac{1}{1+z_L}}^{1}\frac{\dd x}{x^2\sqrt{\Omega_{s}}}\bigg],
\end{align}
while
\begin{align}\label{50b}
 \Omega_{s}=\Omega_\Lambda+\Omega_Kx^{-2}+\Omega_Mx^{-3}+\Omega_Rx^{-4}
\end{align}
\end{subequations}
and
\begin{align}\label{51}
 \Omega_K=\frac{-c^2}{\alpha^2\,a_0^2\,H_0^2}
\end{align}
Also, for the angular diameter distance of the surface of last scattering in spatially closed universe, we have
\begin{align}\label{52}
 d_A=a_Lr_L=\frac{\alpha}{(1+z_L)}\sin\chi_L
\end{align}

In order to extract $C^S_{TE,\ell}$ we need the multipole moment $a^S_{T,\ell m}$ which has been derived in section \ref{604} as
\begin{align}\label{47}
  a^S_{T,\ell m}&=\nonumber\\
  &T_0\sum_{n}\alpha_{\ell m}\bigg[F_n\Pi_{n\ell}(\chi_L)+G_n\dv{\chi_L}\Pi_{n\ell}\bigg]
\end{align}
By combining Eq.(\ref{270}) and Eq.(\ref{47}), we find
\begin{align}\label{48}
C^S_{TE,\ell}&=\frac{3\,T_0^2}{45}\frac{\ell^2\, \bar{t}_\gamma}{a_L r_L^2}\times\nonumber\\
&\sum_{n}\bigg[G_nF_n\Pi_{n\ell}^2(\chi_L)+G_n^2\Pi_{n\ell}\dv{\Pi_{n\ell}(\chi_L)}{\chi_L}\bigg]
\end{align}
The average amount of the second term inside the bracket equals to zero from the WKB approximation. By replacing $F_n$ and $G_n$ from Eqs. (\ref{70}),(\ref{71}) and use the same procedures used for $C^S_{TE,\ell}$ derivation we find that
\begin{align}\label{49}
&\frac{\ell(\ell+1)}{2\pi}C^S_{TE,\ell}=\frac{-4\pi\,T_0^2\,N^2}{125\times\sqrt{3}}\frac{\bar{t}_\gamma}{r_L}(1+z_L)\times\nonumber\\
&\sum_{n}\frac{\ell^3\left(\frac{n/a_0}{\alpha\, k_R}\right)^{n_s-1}}{n^3\sin^3\chi_L\big[\frac{n^2}{\ell^2}-\frac{1}{\sin\chi_L^2}\big]^{1/2}(1+R_L)^{3/4}}\times\nonumber\\
&e^{-\left(\frac{q_nd_D}{a_L}\right)^2}S\left(\frac{q_nd_T}{a_L}\right)\sin\left(\frac{q_nd_H}{a_L}+\Delta\left(\frac{q_nd_T}{a_L}\right)\right)\times\nonumber\\
&\Big[3\,\mathcal{T}\left(\frac{q_nd_T}{a_L}\right)R_L-(1+R_L)^{-\frac{1}{4}}\times\nonumber\\
&e^{-\left(\frac{q_nd_D}{a_L}\right)^2}\mathcal{S}\left(\frac{q_nd_T}{a_L}\right)\cos\left(\frac{q_nd_H}{a_L}+\Delta\left(\frac{q_nd_T}{a_L}\right)\right)\Big]
\end{align}
It is useful now to summarize the dependence of quantities appearing in $C^S_{EE,\ell}$ and $C^S_{TE,\ell}$ on various cosmological parameters. Taking as fixed the value of the present microwave temperature (which gives the values for $\Omega\gamma h^2$ and $\Omega_Rh^2=\Omega\gamma h^2+\Omega\nu h^2$), and fixing the values of $t_L$, $z_L$ and $\sigma$, we see that
\begin{itemize}
  \item $R_L\propto\Omega_B h^2$.
  \item The integral in $d^2_{Silk}$ is complicated but not very sensitive to $\Omega_B h^2$ and $\Omega_M h^2$. Apart from this integral, $d^2_{Silk}$ is proportional to $(\Omega_B h^2)^{-7/2}(\Omega_M h^2)^{-1/2}$.
  \item $d^2_{Landau}$ depends on $\Omega_B h^2$ through the factor $(1+R_L)^{-1}$.
  \item $d_H\propto(\Omega_B h^2)^{-1/2}(\Omega_M h^2)^{-1/2}$ (Aside from a slowly varying logarithm).
  \item $\sin\chi_L$ depends on $H_0, \Omega_\Lambda, \Omega_M, \Omega_B$ and $\Omega_K$.
\end{itemize}
The major advantage of analytic expression is the explicit dependencies on cosmological parameters through above relations. Also, it seems that the closed universe assumption reveals its major effects through hyperspherical Bessel function $\Pi_{n\ell}$ and a sum over $n$ instead of an integral which is appeared in a flat or an open universe assumption.
%%%%%%%%%%%%%%%%%%%%%%%%%%%%%%%%%%%%%%%%%%%%%%%%%%%%%%%%%%%%%%%%%%%%%%%%%%%%%%%%%%%%%%%%%%%%%%%%%%%%%%%%%%%%%%%%%%%%%%%%%%%%%%%%%%%%%%%%%%%%%%%%%%%%%%%%%%%
%Section 5
\section{The EE and TE power spectra; comparison with numerical results and observation}\label{605}
At this section, we plot the scalar mode EE and TE power spectrums derived in the previous section using the latest cosmological parameters and compare the results with CAMB code outputs.
In order to see how well all applied approximations work in practice, we shall calculate $C_{EE,\ell}^S$ and $C_{TE,\ell}^S$ for a realistic set of cosmological parameters from latest observation “Planck 2015”. The cosmological parameters of this set are (TT+Lensing): \cite{r27}
\begin{align*}
  \Omega_Mh^2=0.1416\qquad\Omega_Bh^2=0.02225
\end{align*}
\begin{align*}
  h=0.6781\qquad\Omega_k=1-\Omega_M-\Omega_\Lambda-\Omega_R=-0.005
\end{align*}
and	
\begin{align*}
  n_s=0.9677\qquad k_{\mathcal{R}}=0.05\,Mpc^{-1}
\end{align*}
\begin{align*}
 N^2=\frac{A_s}{4\pi}=1.702\times10^{-10}\qquad \exp(-2\tau_{reion})=0.8763
\end{align*}
We take $T_0=2.725 K$, which yields $\Omega_\gamma h^2=2.47\times10^{-5}$, and by assuming three flavors of massless neutrino we have $\Omega_R h^2=4.15\times10^{-5}$. By studying the thermal history of the universe and recombination process which depends on above cosmological parameters, one also can find the parameters describing recombination as \cite{r24}
\begin{align*}
  1+z_L=1090\quad\sigma_t=262\,K\quad t_L=370,000\,yrs
\end{align*}
By using all above values we find
\begin{align*}
  &R_0=\frac{3\Omega_B}{4\Omega_\gamma}=675.99\\
  &R_L=\frac{3\Omega_B}{4\Omega_\gamma(1+z_L)}=0.6201\\
  &R_{EQ}=\frac{3\Omega_B\Omega_R}{4\Omega_\gamma\Omega_M}=0.1979
\end{align*}
and Eqs.(\ref{42}),(\ref{43}),(\ref{44}),(\ref{52}) gives
\begin{align*}
  &d_A=12.65\,Mpc\qquad d_T=0.1251\,Mpc\\
  &d_H=0.1331\,Mpc\qquad d_D=0.008056\,Mpc
\end{align*}

\begin{figure}
\includegraphics[width=0.45\textwidth]{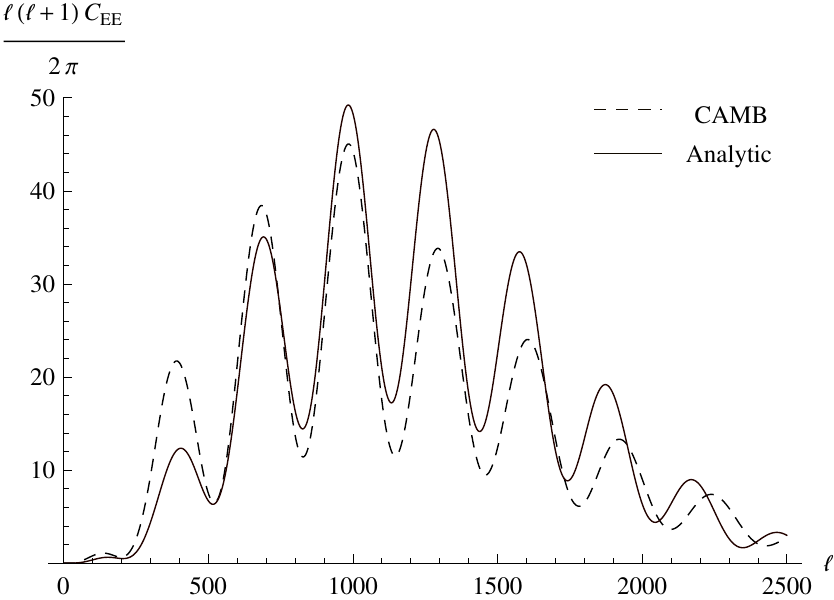}
\caption{The scalar multipole coefficient $\ell(\ell+1)C^S_{EE,\ell}/2\pi$ in square micreoKelvin, vs. $\ell$, for a closed universe with $\Omega_K=-0.005$ in comparison with numerical result from CAMB.}
\label{CL-EE}
\end{figure}

\begin{figure}
\includegraphics[width=0.45\textwidth]{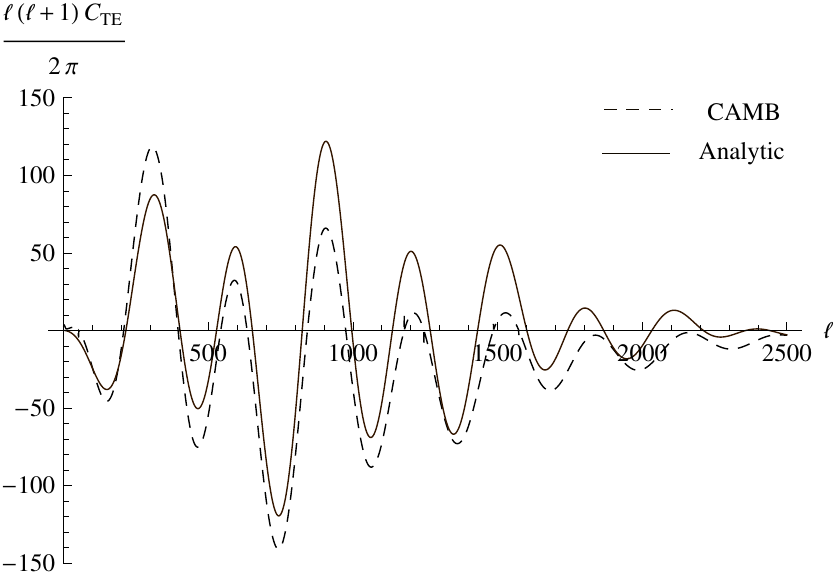}
\caption{The scalar multipole coefficient $\ell(\ell+1)C^S_{TE,\ell}/2\pi$ in square micreoKelvin, vs. $\ell$, for a closed universe with $\Omega_K=-0.005$ in comparison with numerical result from CAMB.}
\label{CL-TE}
\end{figure}

\begin{figure}
\includegraphics[width=0.45\textwidth]{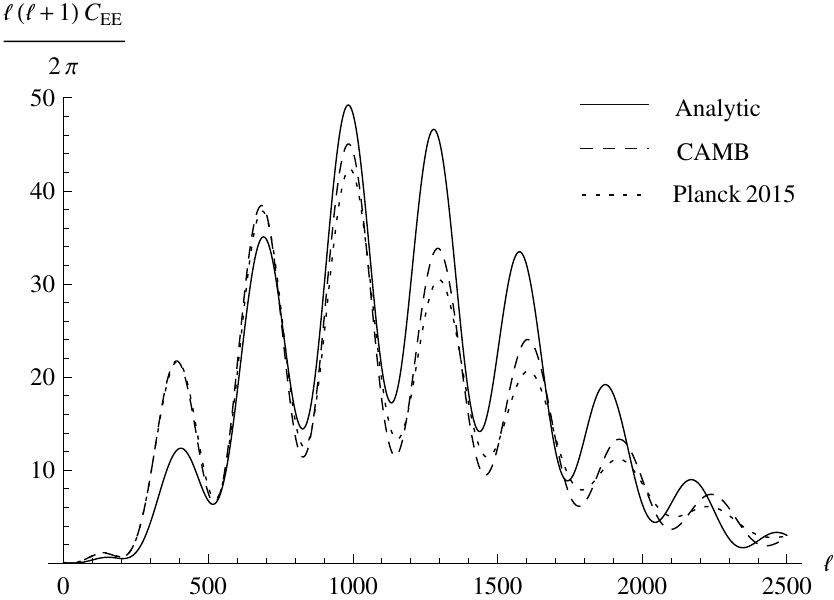}
\caption{The scalar multipole coefficient $\ell(\ell+1)C^S_{EE,\ell}/2\pi$ in square micreoKelvin, vs. $\ell$, for a closed universe with $\Omega_K=-0.005$ in comparison with numerical result from CAMB and Planck 2015 observational result.}
\label{CL-EE-Planck}
\end{figure}

\begin{figure}
\includegraphics[width=0.45\textwidth]{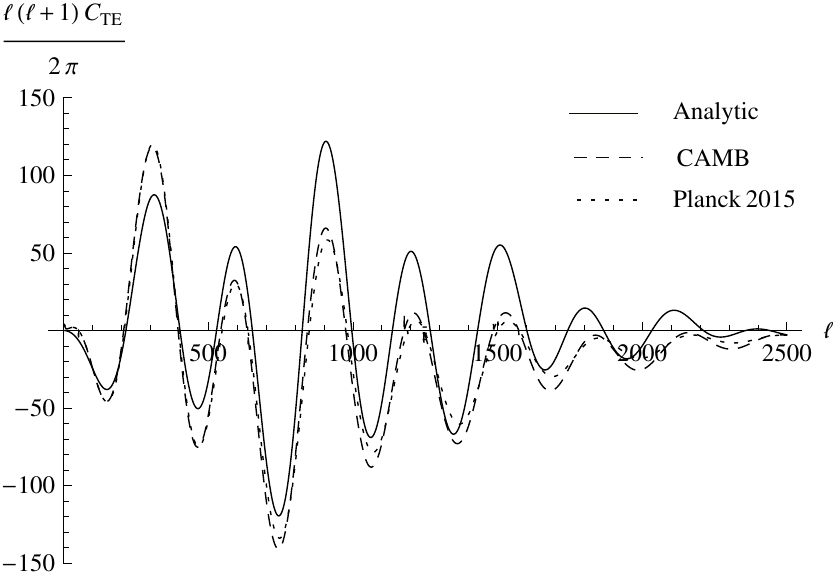}
\caption{The scalar multipole coefficient $\ell(\ell+1)C^S_{TE,\ell}/2\pi$ in square micreoKelvin, vs. $\ell$, for a closed universe with $\Omega_K=-0.005$ in comparison with numerical result from CAMB and Planck 2015 observational result.}
\label{CL-TE-Planck}
\end{figure}

Finally, based on Eqs.(\ref{46}) and (\ref{49}), above parameters and sum over $n$ until a cut-off at $n=3000$ (raising the cut-off more than 3000 has a negligible effect), the scalar multipole coefficient power spectrums in closed universe with $\Omega_k=-0.005$ in comparison with numerical calculation based on the CAMB code are shown in figures \ref{CL-EE} and \ref{CL-TE}.\\
These figures show that the overall profile of the analytic spectra agrees with the numerical result for relatively big and moderate scales until $\ell\simeq1000$. In these scales, the peak positions are in very good agreement with numerical result while the peak heights agree with numerical curve to within $20\%$ due to the approximations have been considered for these derivations (sudden transition from opacity to transparency and take the evolution of perturbations hydrodynamically) \cite{r23}.\\
Just for comparison, the observational results from Planck are also given in figures \ref{CL-EE-Planck} and \ref{CL-TE-Planck}.

\begin{figure}
\includegraphics[width=0.45\textwidth]{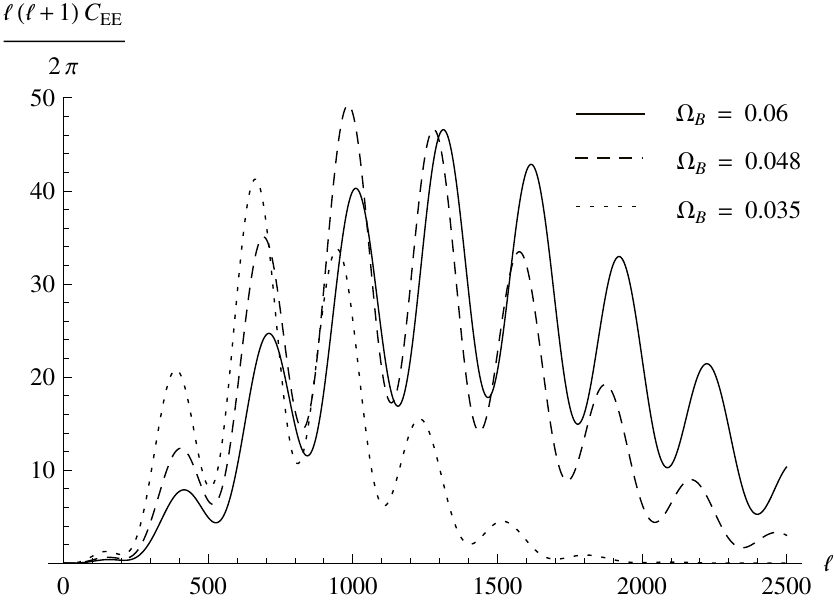}
\caption{Baryon density $\Omega_B$ dependency of the scalar multipole coefficient $\ell(\ell+1)C^S_{EE,\ell}/2\pi$ in square micreoKelvin, vs. $\ell$, for a closed universe with $\Omega_K=-0.005$}
\label{CL-EE-BCompare}
\end{figure}

\begin{figure}
\includegraphics[width=0.45\textwidth]{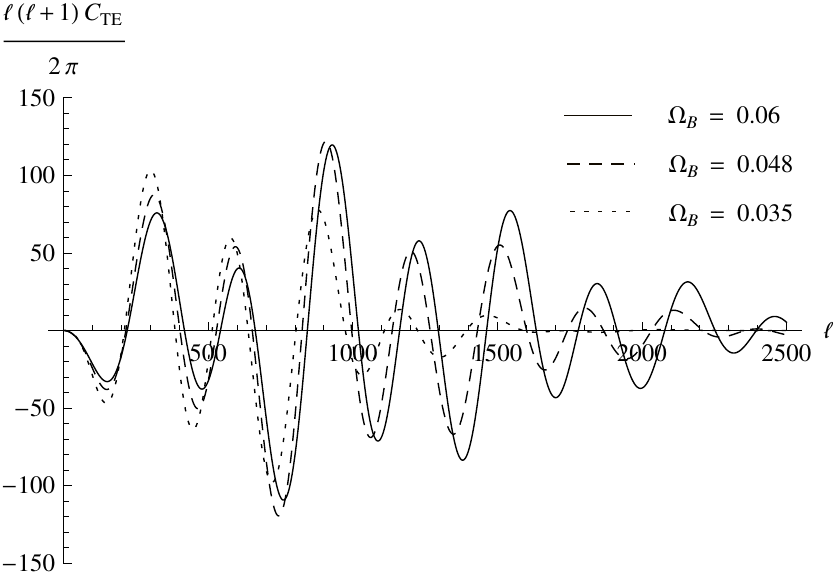}
\caption{Baryon density $\Omega_B$ dependency of the scalar multipole coefficient $\ell(\ell+1)C^S_{TE,\ell}/2\pi$ in square micreoKelvin, vs. $\ell$, for a closed universe with $\Omega_K=-0.005$}
\label{CL-TE-BCompare}
\end{figure}

\begin{figure}
\includegraphics[width=0.45\textwidth]{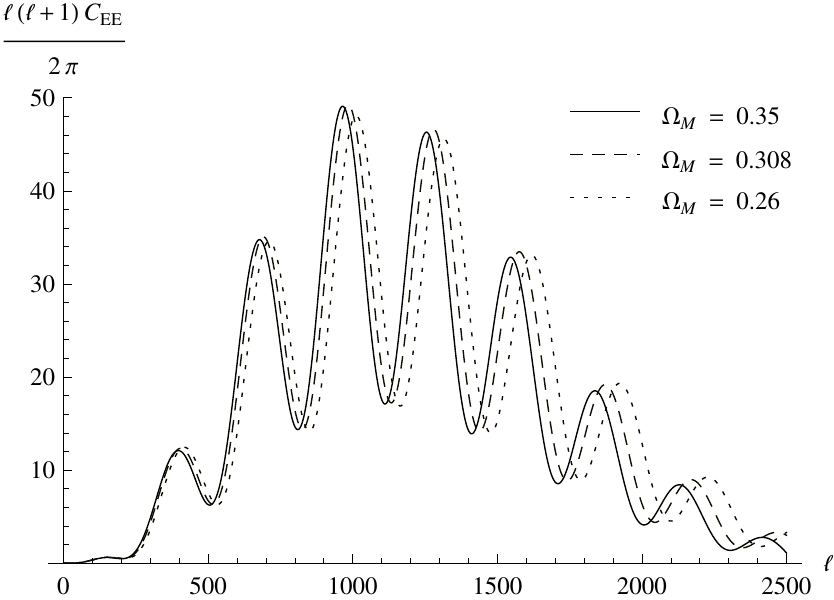}
\caption{Matter density $\Omega_M$ dependency of the scalar multipole coefficient $\ell(\ell+1)C^S_{EE,\ell}/2\pi$ in square micreoKelvin, vs. $\ell$, for a closed universe with $\Omega_K=-0.005$}
\label{CL-EE-MCompare}
\end{figure}

\begin{figure}
\includegraphics[width=0.45\textwidth]{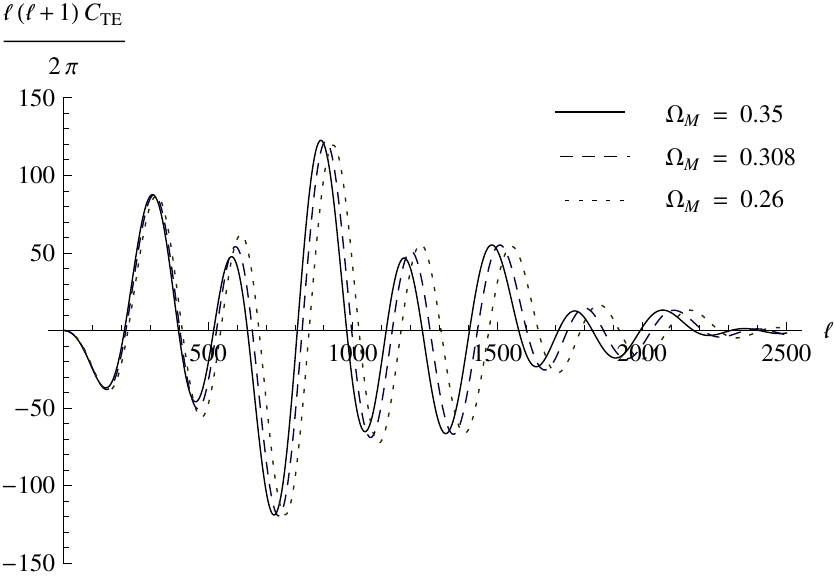}
\caption{Matter density $\Omega_M$ dependency of the scalar multipole coefficient $\ell(\ell+1)C^S_{TE,\ell}/2\pi$ in square micreoKelvin, vs. $\ell$, for a closed universe with $\Omega_K=-0.005$}
\label{CL-TE-MCompare}
\end{figure}

\begin{figure}
\includegraphics[width=0.45\textwidth]{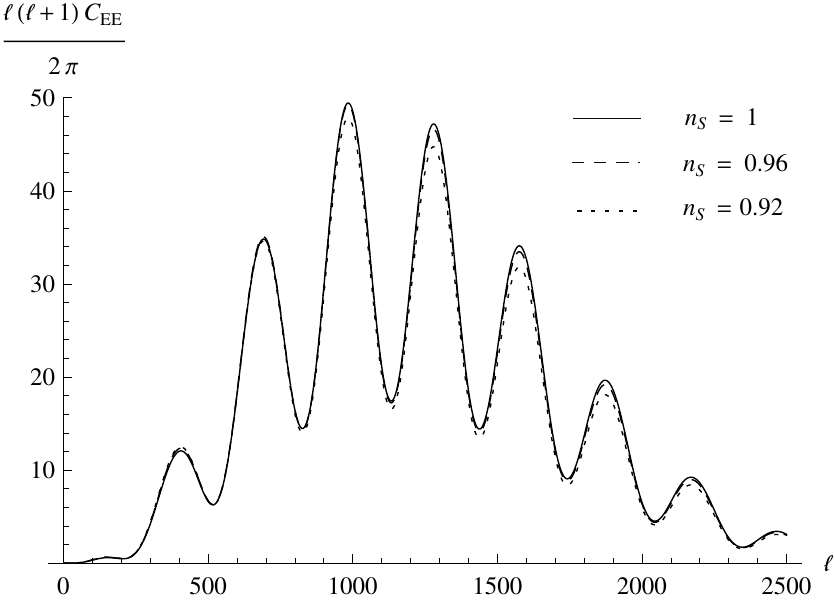}
\caption{Primordial spectral index $n_S$ dependency of the scalar multipole coefficient $\ell(\ell+1)C^S_{EE,\ell}/2\pi$ in square micreoKelvin, vs. $\ell$, for a closed universe with $\Omega_K=-0.005$}
\label{CL-EE-nCompare}
\end{figure}

\begin{figure}
\includegraphics[width=0.45\textwidth]{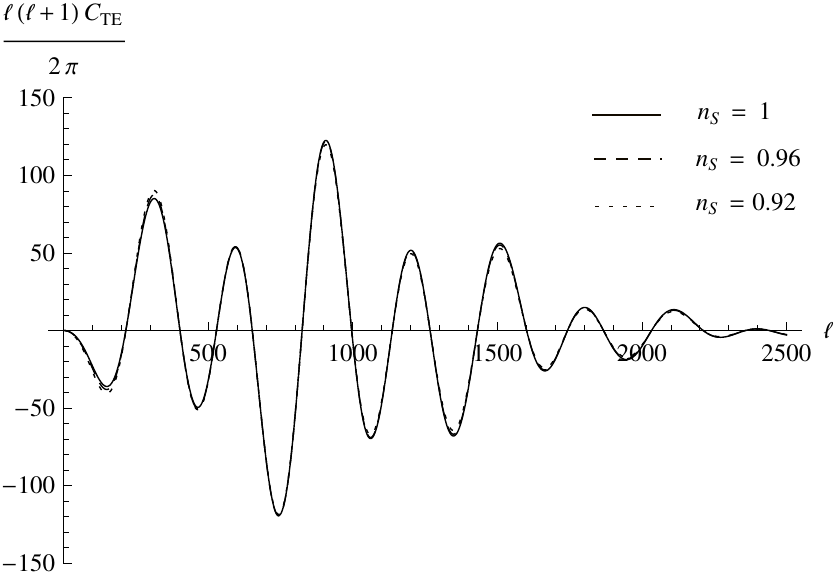}
\caption{Primordial spectral index $n_S$ dependency of the scalar multipole coefficient $\ell(\ell+1)C^S_{TE,\ell}/2\pi$ in square micreoKelvin, vs. $\ell$, for a closed universe with $\Omega_K=-0.005$}
\label{CL-TE-nCompare}
\end{figure}

\begin{figure}
\includegraphics[width=0.45\textwidth]{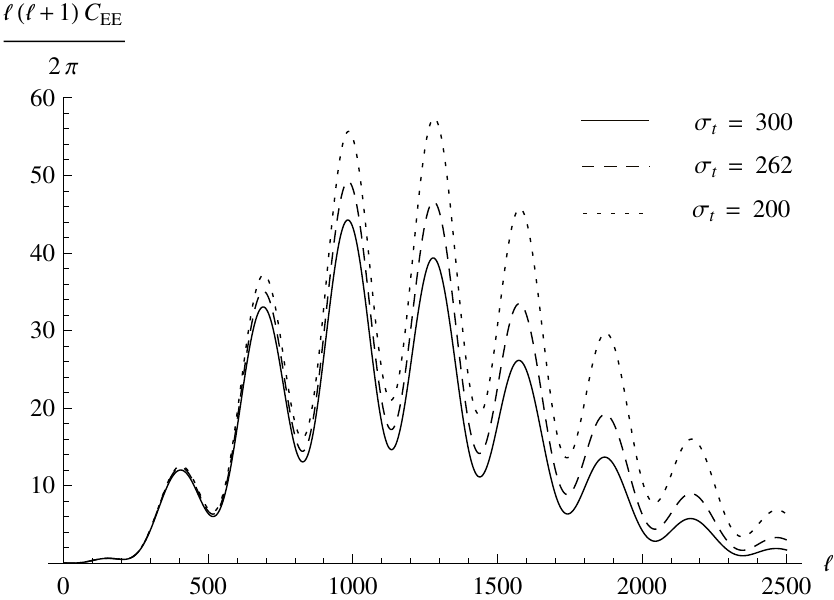}
\caption{Recombination width $\sigma_t$ dependency of the scalar multipole coefficient $\ell(\ell+1)C^S_{EE,\ell}/2\pi$ in square micreoKelvin, vs. $\ell$, for a closed universe with $\Omega_K=-0.005$}
\label{CL-EE-SCompare}
\end{figure}

\begin{figure}
\includegraphics[width=0.45\textwidth]{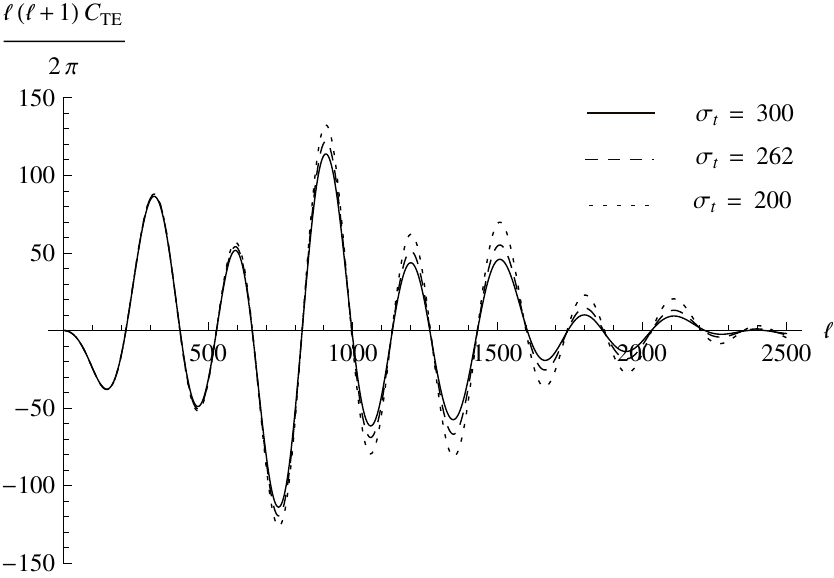}
\caption{Recombination width $\sigma_t$ dependency of the scalar multipole coefficient $\ell(\ell+1)C^S_{TE,\ell}/2\pi$ in square micreoKelvin, vs. $\ell$, for a closed universe with $\Omega_K=-0.005$}
\label{CL-TE-SCompare}
\end{figure}

Figures \ref{CL-EE-BCompare} and \ref{CL-TE-BCompare} shows the dependence of $C^S_{EE,\ell}$ and $C^S_{TE,\ell}$ upon the baryon fraction $\Omega_B$. These figures show that in large scales a greater value of $\Omega_B$ yields lower amplitude of both $EE$ and $TE$ multipole coefficients due to the factor $(1+R_L)$ in denominators which has a direct dependency on $\Omega_B$ through $R_L$. In small scales the damping factor $\exp{-\left(\frac{q_nd_D}{a_L}\right)^2}$ dominates and reduce the amplitudes in both curves. By decreasing $\Omega_B$ this damping is happen in a smaller $\ell$ due to the bigger amount of damping length $d_D$. These figures also show that a greater $\Omega_B$ will shift the location of peaks to larger $\ell$ due to a lower acoustic horizon distance $d_H$.\\
Figures \ref{CL-EE-MCompare} and \ref{CL-TE-MCompare} show that smaller amount of $\Omega_M$ shifts the peaks into higher $\ell$ through the phase shift $\Delta$ in oscillating terms.\\
In figures \ref{CL-EE-BCompare},\ref{CL-TE-BCompare},\ref{CL-EE-MCompare} and \ref{CL-TE-MCompare}, We have changed the amount of $\Omega_\Lambda$ in order to keep $\Omega_K$ unchanged.\\
Figures \ref{CL-EE-nCompare} and \ref{CL-TE-nCompare} show the dependence of $C^S_{EE,\ell}$ and $C^S_{TE,\ell}$ upon the scalar spectral index $n_s$. As it is clear from the figures, $EE$ and $TE$ multipole coefficients do not show a high dependency on the values of $n_s$.\\
Finally, Figures \ref{CL-EE-SCompare} and \ref{CL-TE-SCompare} show that a longer recombination process (a greater $\sigma_t$) yields a lower amplitude of $C^S_{EE,\ell}$ and $C^S_{TE,\ell}$ through damping length $d_D$.\\

%%%%%%%%%%%%%%%%%%%%%%%%%%%%%%%%%%%%%%%%%%%%%%%%%%%%%%%%%%%%%%%%%%%%%%%%%%%%%%%%%%%%%%%%%%%%%%%%%%%%%%%%%%%%%%%%%%%%%%%%%%%%%%%%%%%%%%%%%%%%%%%%%%%%%%%
%Section 6
\section{Conclusion and summary}\label{606}
We have studied the CMB polarization in a model with the spatially closed background. By using some consequences we achieved in our previous work, we have presented a general formula for the scalar mode CMB polarization in a closed universe. By considering some tolerable approximations we have extracted \emph{analytic} formulas for the EE and TE polarization multipole coefficient in a closed universe. These formulas provide a transparent information of CMB polarization and their dependence on the cosmological parameters in a spatially closed background than any other works at the field that gives a generally complicated formula more useful for computer calculations. This achieves by means of some approximations such as assuming a sudden transition from opacity to transparency at the time of last scattering $t_L$, but of course, the drop takes place during some finite interval of time and considering the evolution of perturbation hydrodynamically.\\
We compared the hydrodynamically extracted scalar mode EE and TE polarization power spectra in the closed background with the result of CAMB as a numerical calculation and find a general agreement between the analytic and numeric analysis on large and moderate angular scales. Also, several interesting properties of CMB polarization are revealed by the analytic spectra.\\
As the major advantage of analytic expressions, $C^S_{EE,\ell}$ and $C^S_{TE,\ell}$ explicitly show the dependencies on baryon density $\Omega_B$, matter density $\Omega_M$, curvature $\Omega_K$, primordial spectral index $n_s$, primordial power spectrum amplitude $A_s$, Optical depth $\tau_{reion}$, recombination width $\sigma_t$ and recombination time $t_L$.

%%%%%%%%%%%%%%%%%%%%%%%%%%%%%%%%%%%%%%%%%%%%%%%%%%%%%%%%%%%%%%%%%%%%%%%%%%%
\section*{References}
%\bibliography{mybibfile}

\end{document}